\documentclass{elsart}
\usepackage{graphicx,natbib,amssymb}
\journal{Chemical Physics Letters}
\begin{document}
\begin{frontmatter}
\title{Negative electric field dependence of mobility in TPD doped Polystyrene}
\author[RRCAT]{S. Raj Mohan\corauthref{cor}},
\corauth[cor]{Corresponding author. Tel.: +91 7312488361; Fax: +91 7312488300}
\ead{raj@cat.ernet.in}
\author[RRCAT]{M. P. Joshi}, 
\author[RRCAT]{Manoranjan. P. Singh}
\address[RRCAT]{Laser Physics Applications Division, Raja Ramanna Centre for Advanced Technology, Indore, 
India 452013.}

\begin{abstract}
A total negative field dependence of hole mobility down to low temperature was observed in N,N'-diphenyl-N,N'-bis(3-methylphenyl)-(1,1'-biphenyl)-4,4'diamine (TPD) doped in Polystyrene. The observed field dependence of mobility is explained on the basis of low values of energetic and positional disorder present in the sample. The low value of disorder is attributed to different morphology of the sample due to aggregation/crystallization of TPD. Monte Carlo simulations were also performed to understand the influence of aggregates on charge transport in disordered medium with correlated site energies. The simulation supports our experimental observations and justification on the basis of low values of disorder parameters.\end{abstract}
\end{frontmatter}
\newpage
\section{Introduction}
Films of organic semiconducting materials are widely used in developing various optoelectronic devices like organic light emitting diodes (OLED), organic field effect transistors (OFET), organic solar cells etc[1,2]. Spin cast films of molecularly doped polymers (MDP) are used most widely as active layer in these devices because of simple fabrication technique. These MDP films are mostly amorphous/disordered and have low carrier mobility.  To improve mobility and transport properties various methods, like annealing, irradiation etc. are employed to reduce the disorder in the active layer (i.e. change film morphology)[3-6]. These methods improve structural order in the organic thin film by creating more ordered regions in the otherwise amorphous film. Often, depending upon the processing conditions while film casting, the ordered regions are also formed unintentionally due molecular aggregation / crystallization of dopants in MDP or aggregation of polymer chains in case of conjugated polymers [7,8]. Therefore, in most of the cases where the active layers are MDP or pure polymer, films are not purely disordered or amorphous rather they are partially ordered. The charge transport in the partially ordered films therefore demands deeper understanding of transport mechanism. The assumptions made in hopping based charge transport models, for example the Gaussian Disorder Model(GDM) [9], developed for a completely isotropic and disordered medium  probably not sufficient [10]. Moreover these models of charge transport do not consider morphology of the active layers in finer details. The presence of ordered regions can in-fact reduce the overall energetic disorder in the material [5,7] and that enhances the mobility but substantially influence the charge transport mechanism [7,11,12]. Field and temperature dependence of mobility in such partially ordered samples show wide variation with morphology of the sample and is subject of intense research. In general the field dependence of the mobility in polymer films follows Poole-Frenkel type behavior. However, in certain cases, at low temperature, either very weak field dependence or even negative field dependence of mobility has also been reported [13-16]. Reports also suggest dependence of mobility on temperature as $1/T$ as well as $1/T^2$ dependence. In some cases it is hard to distinguish between the two power law behavior of the mobility [13,14,17].
 
In Poole-Frenkel type behavior, the mobility increases with increase of electric field in a $\log\mu Vs. E^{1/2}$ fashion. According to GDM the increase of mobility in $\log\mu Vs. E^{1/2}$ fashion in disordered molecular solids is due to the tilting of density of states by the applied potential that lead to the decrease of energetic barrier as seen by the charge carriers in its transit [9]. At higher electric field strength the energetic barrier seen by the carrier is negligibly small and this results in the saturation of drift velocity of the carrier. Once the drift velocity get saturated the mobility decrease with further increase of electric field, i.e. negative field dependence of mobility[18].  At intermediate field strength when the temperature is low, the field dependence of mobility remains positive. When temperature increases carrier gains more thermal energy to over come the energetic barrier. Thus the mobility increases and the field dependence of mobility weakens. At higher temperature energetic barrier seen by the carrier is negligibly small which results in negative field dependence of mobility even at intermediate field strengths. Lower the energetic disorder inside the sample lower is the temperature at which negative field dependence can be observed [7a,11,13b]. In principle if disorder is very low one can observe negative field dependence of mobility at lower temperature. The above reasoning is also justified on the basis of GDM. GDM predicts the variation of the mobility as given by equation[9], 
\begin{equation}
\mu=\mu_0exp\left(-\left[\frac{2\sigma}{3kT}\right]^2\right)exp\left[C\left\{\left(\frac{\sigma}{kT}\\\right)^2-\Sigma^2\right\}E^{1/2}\right]
\end{equation}
where $\mu_{0}$ is a prefactor mobility, $\sigma$ is the measure of energetic disorder (width of the Gaussian distribution of site energies), $\Sigma$ is the measure of positional disorder, the measure of geometrical disorder, $k$ is the Boltzmann constant, $T$ is the temperature in Kelvin and \emph{C} is an empirical constant. From Equation (1), the term $ \left\{(\frac{\sigma}{kT})^2-\Sigma^2\right\}$ decides the slope of field dependence of mobility at the intermediate field regime, where the mobility shows $\log\mu Vs. E^{1/2}$ dependence.  When temperature increases the term  becomes negative. Also, if the value of $\sigma$ is low then the above term become negative at lower temperatures. This suggests that if the overall energetic disorder of the film is very low or if the charge transport occurs through regions of very low energetic disorder then the slope of field dependence of mobility can remain negative down to lower temperatures. Similarly if $(\frac{\sigma}{kT})^2<\Sigma^2$, the mobility shows a negative field dependence. This generally happens only when the positional disorder in the sample is remarkably very high. It has also been shown using Monte Carlo simulations that high positional disorder can lead to negative field dependence of mobility at lower electric field strength [9]. In MDPs the negative field dependence has generally been observed at very high temperature and also when the concentration of the dopant is very low [3]. 

In this paper we show negative field dependence of mobility down to low temperatures in MDP. We have investigated the field and temperature dependence of mobility in films of N,N'-diphenyl-N,N'-bis(3-methylphenyl)-(1,1'-biphenyl)-4,4'diamine (TPD) dye.  TPD dye is a well known blue emitting laser dye and also a hole transporting dye. Films of TPD have been used widely in fabricating various organic devices [2,3]. We used TPD doped in  Polystyrene (PS) at 40:60 proportions by weight (TPD:PS) and measured hole mobility using time-of-flight(TOF) transient photoconductivity technique. A total negative field dependence of mobility was observed down to low temperature $\sim$150K. This was attributed to the low value of disorder in the sample due to the presence of aggregation/crystallization of TPD molecules. The study highlights the drastic change in morphology of the sample and the resulting reduction in the overall disorder of a molecularly doped polymer upon aggregation/crystallization of dopant. Monte Carlo simulation was performed to understand the influence of aggregates on charge transport with correlated site energies. Simulation supports our experimental observation as well as the justification provided on the basis of low disorder present in the sample.  

\section{Experimental and Simulation Details}
\subsection{Experimental details}
Solutions of TPD doped Polystyrene at 40:60 proportions by weight were made by dissolving the required amount of TPD and PS in Chloroform. Films were spin-cast on to a neatly cleaned fluorine doped tin oxide (FTO) coated glass substrate.  Samples were kept in vacuum for 24 hours at ambient temperature to remove the residual solvent. A thin layer of amorphous Selenium (a-Se) and an Aluminum top electrode was deposited on to it by thermal evaporation. All coatings were done at base pressure of 10$^{-5}$mbar.  Thickness and capacitance of the samples were $\sim$4$\mu$m and $\sim$10pF respectively. 
Field and temperature dependence of mobility in these samples was determined using conventional small signal Time of Flight (TOF) transient photoconductivity technique [3]. Samples were mounted on a homemade cryostat to perform temperature dependent studies. A variable DC potential was applied across the device such that no injection occurs from the electrodes to the sample. A 15ns laser pulse from second harmonic of Nd: YAG laser (532nm) was used for generating a thin sheet of charge in a-Se at a-Se/ TPD:PS  interface. Laser intensity is adjusted so that total charge generated is less than 0.05\emph{CV}, where \emph{C} is the capacitance of the device and \emph{V} is the voltage applied across the device. The time resolved photocurrent is acquired using an oscilloscope as a voltage across a load resistance. The transit time, $\tau$, is obtained from photocurrent signal and the mobility is calculated using  $\mu= L^2/V\tau$ , where $L$ is the thickness of the sample. Morphology of the samples was characterized using Scanning electron microscope (SEM) images.

\subsection{Details of Monte Carlo Simulation}

Monte Carlo simulations wereperformed to support our experimental observations and   explore the influence of aggregates on field and temperature dependence of mobility.  The Monte Carlo simulation is based on the commonly used algorithm reported by Sch\"{o}nherr et al [19].  A lattice of 70x70x70, along x, y and z direction, with lattice constant a = 6\r{A} was used for computation. Z direction is considered as the direction of the applied field. The size of the lattice is judged by taking into account of the available computational resources. The site energies of lattice were initially taken randomly from a Gaussian distribution of mean $\sim$5.1eV and standard deviation $\sigma$ = 75meV, which gives the energetic disorder parameter $\hat{\sigma}$=$\sigma/kT$ . The value for $\sigma$ was chosen close to the experimental value observed in TPD based MDPs [3]. The site energies were made correlated by considering the energy of a site as an average of energies of neighboring sites which is defined as follows [20],

\begin{equation}
\varepsilon_{i}=N^{-1/2}\sum_{j} K(r_{ij})\varepsilon_{j}
\end{equation}
Where the variable $\varepsilon_{j}$ denotes the uncorrelated energies on the neighboring sites, \emph{N} is the normalization factor that results in required standard deviation and \emph{K} is kernel that provides a degree of correlation among the sites [20]. In our simulation, kernel \emph{K} is considered as unity within a sphere of radius \emph{a} (\emph{a} is the intersite distance) and zero outside this sphere. Simulation was performed on this energetically disordered lattice with the assumption that the hopping among the lattice sites was controlled by Miller-Abrahams equation [21] in which the jump rate  $ \nu_{ij}$ of the charge carrier from the site \emph{i} to site \emph{j} is given by

\begin{equation}
\nu _{ij}  = \nu _0 \exp \left( { - 2\gamma a\frac{{\Delta R_{ij} }}{a}} \right)\left( {\begin{array}{*{20}c}
   {\exp \left( { - \frac{{\varepsilon _i  - \varepsilon _j }}{{kT}}} \right)\exp \left( { \pm \frac{{eEa}}{{kT}}} \right)} & { \to \varepsilon _j  > \varepsilon _i }  \\
   1 & { \to \varepsilon _j  < \varepsilon _i }  \\
\end{array}} \right)
\end{equation}

where $E$ is the applied electric field, \emph{a} is the intersite distance, $k$ is the Boltzmann constant, $T$ is the temperature in Kelvin, $\Delta R_{ij}=R_{i}-R_{j}$   is the distance between sites \emph{i} and \emph{j} and $2\gamma a$ is the wave function overlap parameter which controls the electronic exchange interaction between sites. Throughout the simulation we assume $2\gamma a=10$[9,19].

Film morphology was varied by incorporating cuboids of so called ordered regions (representing molecular aggregates/microcrystallites in MDPs) of varying size that are placed randomly inside the otherwise disordered host lattice.  Sizes of ordered regions were limited to a maximum size of 25x25x40 sites along x, y and z directions. The energetic disorder inside the ordered region was kept low compared to the lattice. This is justified because the aggregates are more ordered regions and hence to simulate the charge transport the cuboids must be of low energetic disorder compared to host lattice. The site energies inside the ordered regions were also taken randomly from another Gaussian distribution of standard deviation $\sim$15meV (we chose 5 times less compared to host lattice). Earlier reports have even suggested a ten fold reduction of energetic disorder inside the aggregates [22]. Further details of simulation with varying film morphology are provided in Ref [11]. The site energy of the ordered region was also correlated as explained above. The mean energies of ordered regions were chosen such that their difference from the mean energy of host lattice is in the order of \emph{kT}. This is justified by the fact that the aggregation of dopants can also lead to change in the energy gap (shift in HOMO, LUMO levels) and hence the mean energy of Gaussian distribution. Simulations were performed by varying the concentration of such ordered regions (varying the percentage of volume of lattice occupied by ordered region), temperature and electric field so as to simulate the field and temperature dependence of mobility.

\section{Results and Discussions}
\subsection{Experimental Results}
Fig.1 shows the typical time of flight transient signal obtained for TPD:PS (40:60 wt\%) at 290K. Transient showed some plateau suggesting non-dispersive transport. We observe dispersive transport at low electric field strength and low temperature. The transit time was always determined from the double logarithmic plot as shown in the inset of Fig. 1.  Fig. 2 shows the field dependence of mobility parametric with temperature. Negative field dependence of mobility was observed down to low temperature and through out the range of electric field studied. The mobility becomes almost field independent at 150K. There is no remarkable increase in mobility above room temperature compared to lower temperatures. Data at low electric field strength, at low temperature, were not recorded due to very low magnitude of photocurrent transient signal. At high temperatures, in the low field regime, a slight increase in mobility was observed with decrease of electric field. Temperature dependence of zero field mobility and the slope ($\beta$) of intermediate field region of $\log\mu(E=0) Vs. E^{1/2}$ follow $T^{-2}$ as predicted by GDM (Figure 2(b) and 2(c) respectively) [9]. Based on GDM formalism we estimated the disorder parameters. The obtained values of $\sigma,\Sigma$ and $\mu_0$ are 0.0287eV, 2.277 and 2.8471x10$^{-2}$ cm$^2$/Vs respectively.  The value of energetic disorder and positional disorder in our sample is low compared to the reported values for similar and other molecularly doped polymer systems [3,9]. Hence the observed negative field dependence of mobility down to low temperature ($\sim$150K) in TPD:PS  can be attributed to the presence of very low value of energetic disorder in the sample studied. As explained above (see introduction), low of value of energetic disorder can lead to negative field dependence of mobility down to low temperature. When the disorder is very low then the barrier offered to carrier is also very low. Thus drift velocity of the carrier saturates at lower electric field strength which lead to the decrease of mobility with increases of field strength [9,16].

The morphology of TPD:PS in our study was investigated using SEM images. Fig. 3 shows the SEM images of TPD:PS in our study. SEM images showed that TPD has undergone aggregation. Aggregation of TPD resulted in the formation of crystalline regions of few micron sizes and also chains of such microcrystals that are spread all over the sample. Earlier reports of crystallization/aggregation of TPD in TPD doped Polystyrene system, even at low concentration and without annealing[23], also supports our observation. Earlier reports of mobility measurement, who apparently report only positive field dependence, in TPD doped polymers assert that their study was performed in totally amorphous sample with no signs of aggregation of dopants at least during the time of experiment [2,3,24].  Presence of more ordered regions can drastically change the entire morphology of the sample and can lead to reduction in the effective energetic and positional disorder. This is consistent with our observation of low energetic and positional disorder in TPD:PS. Due to aggregation the charge transport therefore occurs through a combination of ordered and less ordered regions. In such cases the charge transport is highly influenced by the packing and orientation of ordered regions. If the charge transport occurs mostly through these ordered regions, regions of very low disorder, then a negative field dependence of mobility down to low temperature can be expected. So the presence of these microcrystallites can drastically change the morphology with lower energetic and positional disorder and the behavior of charge transport in these samples.

\subsection{Simulation Results}

In order to justify and support the above explanation a Monte Carlo simulation of charge transport in disordered lattice was also performed. Aim of the simulation was to understand the influence of aggregates on charge transport, in particular the negative field dependence. Simulated field and temperature dependence of mobility for a pure host lattice having DOS with standard deviation $\sim$75meV (without considering any ordered regions) has been discussed in detail in our earlier report [11]. Simulation results were as predicted by GDM. To study the influence of embedded micro crystals/aggregation of dopants on charge transport, the simulation was performed after incorporating ordered regions inside the host lattice (as explained in simulation procedure above). Fig. 4 shows the field dependence of mobility, at $\sim$248K, parametric with the concentration of ordered regions having DOS with standard deviation $\sim$15meV and mean energy lower by $\sim$ kT compared to mean energy of host lattice. Magnitude of mobility at all regimes of electric field increases with increase of concentration of ordered regions concomitant with decrease of slope at the intermediate field regime. The saturation of mobility and decrease of mobility with further increase of electric field, at high field regime, was observed when the concentration of ordered region was higher than 60\%. With increase in concentration of ordered region the saturation of mobility occurs at lower electric field strength. For the cases when the concentration of ordered region is less that 60\% the mobility was not completely saturated even for the maximum strength of electric field used for simulation. When the concentration of the ordered regions inside the host lattice is very high ($\sim$98\%) the field dependence of mobility become totally negative even at this low temperature ($\sim$248K). The mobility decreases with increase of electric field right from low electric field strengths used for simulation. 

The observed features in the field dependence of the mobility, after incorporating the ordered regions inside the host lattice, can be explained on the basis of decrease of effective energetic disorder in the host lattice when ordered regions are embedded in it. If the energetic disorder is small then the energetic barrier seen by the carriers due to disorder will be small. This results in higher mobility but weaker field dependence. This explains the increase in magnitude of mobility and decrease in the slope of $semi\log\mu Vs. E^{1/2}$ curve at intermediate field regime with increase in concentration of ordered regions in the host lattice. The low value of energetic disorder also can lead to saturation of drift velocity/mobility at lower electric field strength. When the concentration of ordered regions inside the host lattice is very high, the overall energetic disorder will be very low. At such high concentration the carrier will travel mostly through ordered region only. Since the energetic disorder inside the ordered region is small (only $\sim$15meV in our study) the mobility saturates at low electric field strength and even at low temperatures. In such a case one can observe total negative field dependence of mobility down to low temperatures. Moreover for a very low value of energetic disorder the term $ \left\{(\frac{\sigma}{kT})^2-\Sigma^2\right\}$ remains negative down to low temperature. It is also possible to have total negative field dependence down to low temperature when the charge transport occurs totally through ordered regions, i.e. when the whole host lattice is completely occupied with ordered region (Fig. 4, 100\%). Fig. 5 shows the field dependence of mobility parametric with temperature when host lattice is completely occupied with ordered region. A total negative field dependence of mobility down to low temperature is observed but the mobility in this case decreases with increase of temperature. Higher magnitude of mobility was observed for lower temperature ($\sim$150K) and low magnitude of mobility at higher temperature ($\sim$390K). This kind of behavior is possible when the overall energetic disorder is very low. When energetic disorder is very low and the thermal energy is comparable to energetic disorder then thermal energy dominates which forces the carriers to move in longer paths. Effectively the transit time increases and mobility decreases with increase of temperature [16].       
       
From the simulation it is inferred that charge transport in TPD:PS occurs mostly through aggregates but not completely through aggregates. Hence the effective disorder seen by the carrier is low which results in total negative field dependence of mobility. Experimentally we observed that mobility decreases with decreases of temperature which rules out the possibility that charge transport is occurring completely through aggregates. The influence of small disordered regions present in entire trajectory of charge transport (major part of the trajectory is occupied by ordered region) for each carrier become prominent only at low temperature. When the thermal energy of carrier is low, the carrier will see the effect of disorder and that results in the positive field dependence of mobility at low temperature. Fig. 6(a) shows the field dependence of mobility parametric with temperature when 98\% of the  host lattice is occupied with ordered regions having DOS with standard deviation $\sim$15meV and mean energy lower by $\sim$ kT compared to mean energy of host lattice. Mobility decreases with decrease of temperature and positive field dependence of mobility was observed at lower temperature (200K). Negative field dependence of mobility was observed down to 248K. At temperature around 300K, there is no variation of mobility with increase of temperature for all electric field strength used for simulation. This observation is similar to what observed experimentally in TPD:PS at higher temperatures. The temperature dependence of zero field mobility and slope of intermediate field region of log  versus $E^{1/2}$ plot followed $1/T^2$ dependence as predicted by GDM (Data not shown). 

\section{Conclusion}
	Time of flight mobility measurement was carried out in TPD:PS. A total negative field dependence of mobility down to low temperature was observed. This was explained on the basis of low value of energetic disorder in TPD:PS. The low value of energetic disorder was attributed to aggregation of TPD that results in the formation of crystalline regions of few micron size and chaining of such crystals. The presence of aggregates drastically changes the morphology of the sample and reduces the overall disorder in the sample. Monte Carlo simulation study clearly showed the influence of aggregates on charge transport which supports our experimental observation. The simulation suggests if charge transport occurs mostly through aggregates (crystalline regions of very low energetic disorder) the mobility saturates at low electric field strengths and it can lead to a total negative field dependence of mobility down to low temperature. From the simulation studies it is inferred that the charge transport in TPD:PS in the present study are not occurring totally through aggregates but in combination of aggregates and disordered medium. Thus our experimental and simulation studies showed the influence of morphology of the sample on field and temperature dependence of mobility. The study also highlights the need to consider film morphology in various models used for analyzing charge transport in polymer films.

\newpage
\section{References}
\begin{enumerate}
\item	J. Shinar, V. Savvateev, in: J. Shinar (Ed.) Organic Light Emitting Devices: A Survey, Springer-Verlag, New York, 2004.
\item	G. Hadziioannou, Paul F.van Hutten, Semiconducting Polymers: Chemistry, Physics and Engineering, Wiley-VCH, Weinhelm, 2000.
\item	P. M. Borsenberger and D. S. Weiss, Organic Photoreceptors for Xerography,Vol. 59 of Optical engineering series, Marcel Dekker, New York, 1998.
\item	H. Sirringhaus, P. J. Brown, R. H. Friend, M. M. Nielsen, K. Bechgaard, B. M. W. Langeveld-Voss, A. J. H. Splering, R. A. J. Janssen, E. W. Meijer, P. Herwig, D. M. de Leeuw, Nature  401 (1999) 685.
\item	 A Kadashchuk, A. Andreev, H. Sitter, N. S. Sariciftci, Y. Skryshevski, Adv. Funct. Mater. 14 (2004) 970
\item	S. Cho, K. Lee, J. Yuen, G. Wang, D. Moses, A. J. Heeger, M. Surin, R. Lazzaroni, J. Appl. Phys. 100 (2006) 114503.
\item	(a) S. Raj Mohan, M. P. Joshi, Solid State. Commun. 139 (2006) 181. (b) S. Raj Mohan, M. P. Joshi, A. K. Srivastava, Synth. Met., 155, 372 (2005); 
\item	(a) Thuc-Quyen Nguyen, Ignacio B. Martini, Jei Liu, Benjamin J. Schwartz, J. Phys. Chem. B, 104 (2000) 237. (b) H. Aziz, Z. Popovic, S. Xie, Ah-Mee Hor, Nan-Xing Hu,C. Tripp,G. Xu, Appl. Phys. Lett., 72 (1998) 756.
\item	(a) H. B\"{a}ssler, Phys. Stat. Sol. (b), 175 (1993) 15 (b) P. M. Borsenberger, L. Pautmeier, H. B\"{a}ssler, J. Chem. Phys. 94 (1991) 5447 (c) L. Pautmeier, R. Richert, H. B\"{a}ssler, Synth. Met. 37 (1990) 271.
\item	 M. Pope, C. E. Swenberg, Electronic Processes in Organic Crystals and Polymers, Oxford University Press, NewYork, 1999.
\item	 S. Raj Mohan, M. P. Joshi, M. P. Singh, Org. Electron, 9 (2008), 355.
\item	  A. R. Inigo, Hsiang-Chih Chiu, W. Fann, Ying-Sheng Huang, U-Ser Jeng, Tsang-Lang Lin, Chia-Hung Hsu, Kang-Yung Peng, Show-An Chen, Phys. Rev. B. 69 (2004) 075201
\item	  (a) A. J. Mozer, P. Denk, M. C. Scharber, H. Neugebauer, N. S. Sariciftici, P. Wagner, L.  Lutsen, D. Vanderzande, J. Phys. Chem. B 108 (2004) 5235. (b) A. J. Mozer, N. S. Sariciftci, Chem. Phys. Lett. 389 (2004) 438
\item	D. Hertel, H. B\"{a}ssler, U. Scherf, H. H. Hörhold, J. Chem. Phys. 110 (1999) 9214. 
\item	T. Kreouzis, D. Poplavskyy, S. M. Tuladhar, M. Campoy-Quiles, J. Nelson, A. J. Campbell, and D. D. C. Bradley, Phys. Rev. B 73 (2006) 235201.
\item	 S. J. Matrin, A. Kambili, A. B. Walker, Phys. Rev. B 67 (2003) 165214.
\item	S. V. Rakhmanova, E. M. Conwell, Appl. Phys. Lett. 76 (2000) 3822.
\item	 Y. N. Gartstein, E. M. Conwell, J. Chem. Phys. 100 (1994) 9175.
\item	 G. Sch\"{o}nherr, H. B\"{a}ssler, M. Silver, Philos. Magz. 44 (1981) 47.
\item	Yu. N. Gartstein and E. M. Conwell, Chem. Phys. Lett., 245 (1995) 351.
\item	A. Miller, E. Abrahams, Phys. Rev. B 120 (1960) 745 
\item	 J. R. Durrant, J. Knoester, D. A. Wiersma, Chem. Phys. Lett. 222 (1994) 450. 
\item	S. Heun, P. M. Borsenberger,  Physica B 216 (1995) 43. (b)  J. L. Maldonado, M. Bishop, C. Fuentes-Hernandez, P. Caron, B. Domercq, Y. D. Zhang, S. Barlow, S. Thayumanavan, M. Malagoli, J. L. Br\'edas, S. R. Marder, and B. Kippelen, Chem. Mater., 15 (2003) 994

\end{enumerate}

\newpage
\begin{center}
\title{Figure Captions} 
\end{center}

\textbf{Figure 1.}	Typical time of flight transient signal for TPD:PS at 4x10$^5$V/cm, 290K. Inset show	the double logarithmic plot used for determining the transit time.\newline
\textbf{Figure 2.}	Experimental data for TPD:PS(40:60) (a) Electric field dependence of mobility $semi\log\mu Vs. E^{1/2}$ parametric with temperature. (b) $semi\log\mu(E=0)$ Vs. $T^{-2}$  (c)$ \beta$ Vs. $(\sigma/kT)^2$ plot,  where $\beta$ is the slope of high field region of $semi\log\mu Vs. E^{1/2}$ plot.\newline
\textbf{Figure 3.}	 SEM image of TPD:PS showing crystallization/aggregation of TPD molecules.\newline
\textbf{Figure 4.}	Simulated field dependence of mobility for host lattice ($\sigma$=75meV), at 248K, 	embedded with ordered regions of various concentrations having DOS of standard 	deviation$\sim$15meV.\newline
\textbf{Figure 5.}	Field dependence of mobility of host lattice with $\sim$15meV parametric with	temperature. Case is similar when host lattice is completely occupied with ordered	regions.\newline
\textbf{Figure 6.}	Field dependence of mobility of host lattice, parametric with temperature, with 
embedded ordered regions having DOS with standard deviation $\sim$15meV. Concentration of embedded ordered region inside the host lattice is $\sim$98\%. Straight line show linear fit to intermediate field region.

\newpage
\newpage
\begin{figure}
\includegraphics*[width=15cm]{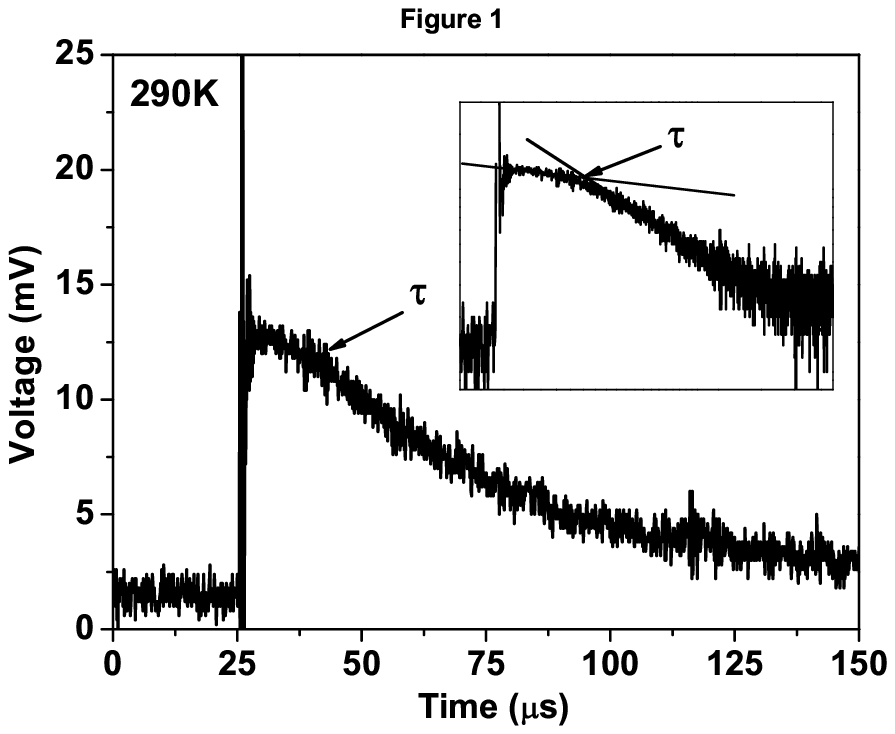}

\label{Fig1}
\end{figure}

\newpage
\begin{figure}
\includegraphics*[width=15cm]{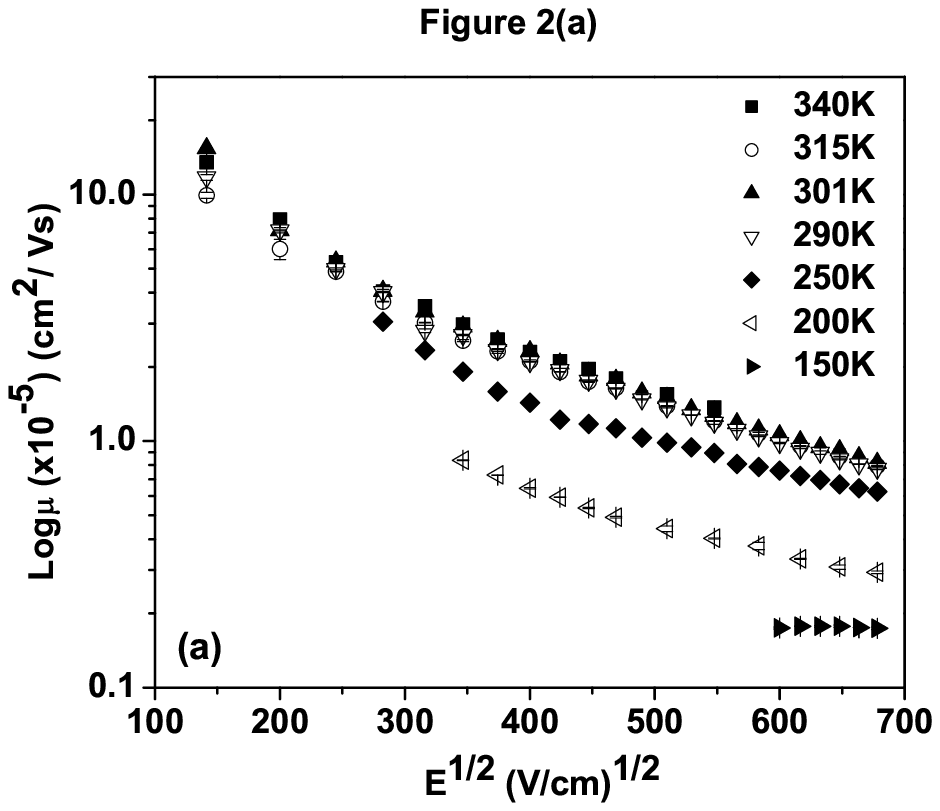}
\label{Fig2a}
\end{figure}
\newpage
\begin{figure}
\includegraphics*[width=15cm]{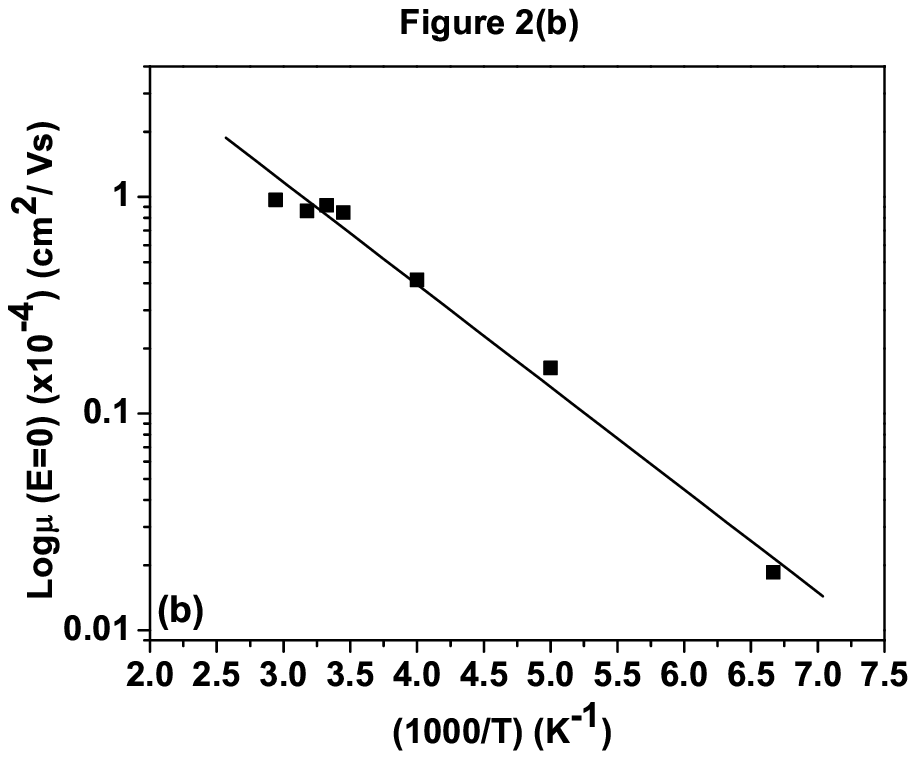}

\label{Fig2b}
\end{figure}

\newpage
\begin{figure}
\includegraphics*[width=15cm]{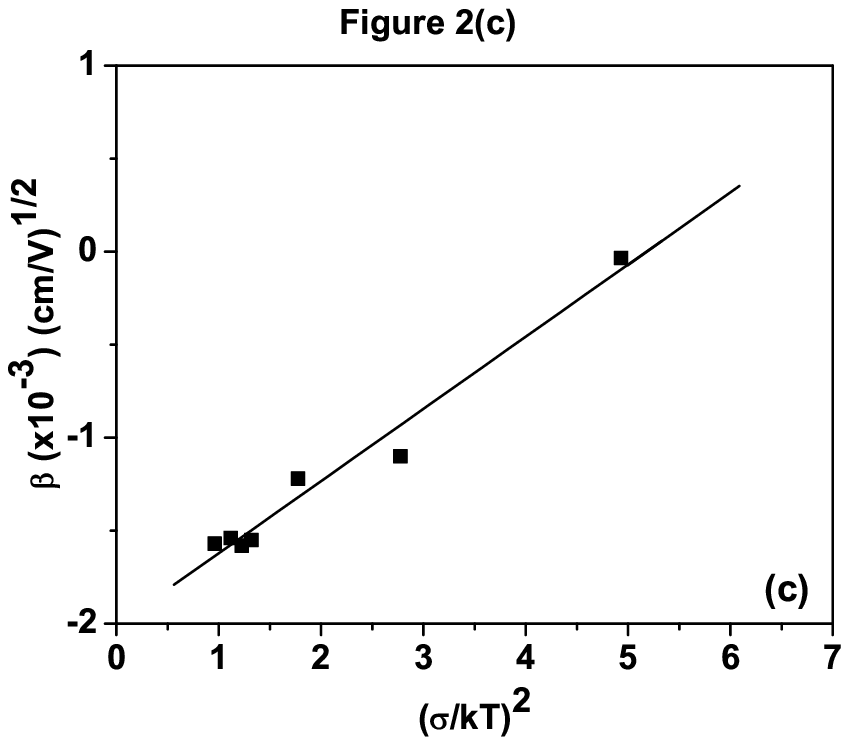}

\label{Fig2c}
\end{figure}

\newpage
\begin{figure}
\includegraphics*[width=15cm]{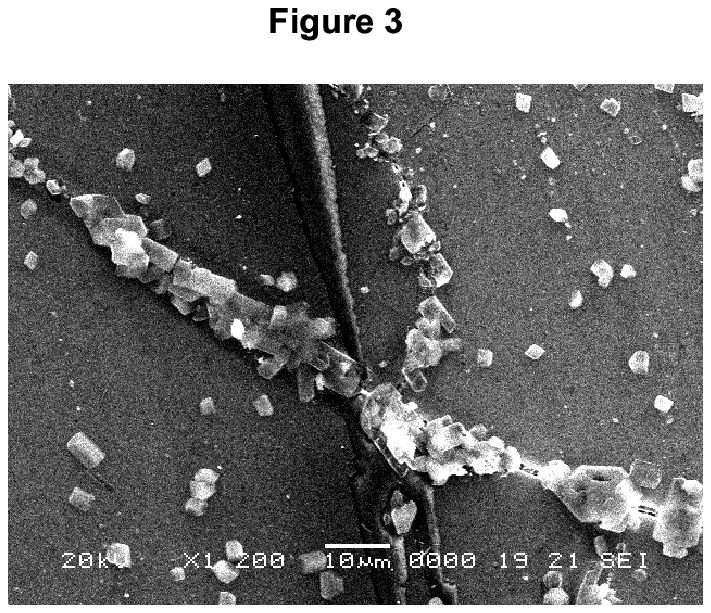}
\label{Fig3}
\end{figure}

\newpage

\begin{figure}
\includegraphics*[width=15cm]{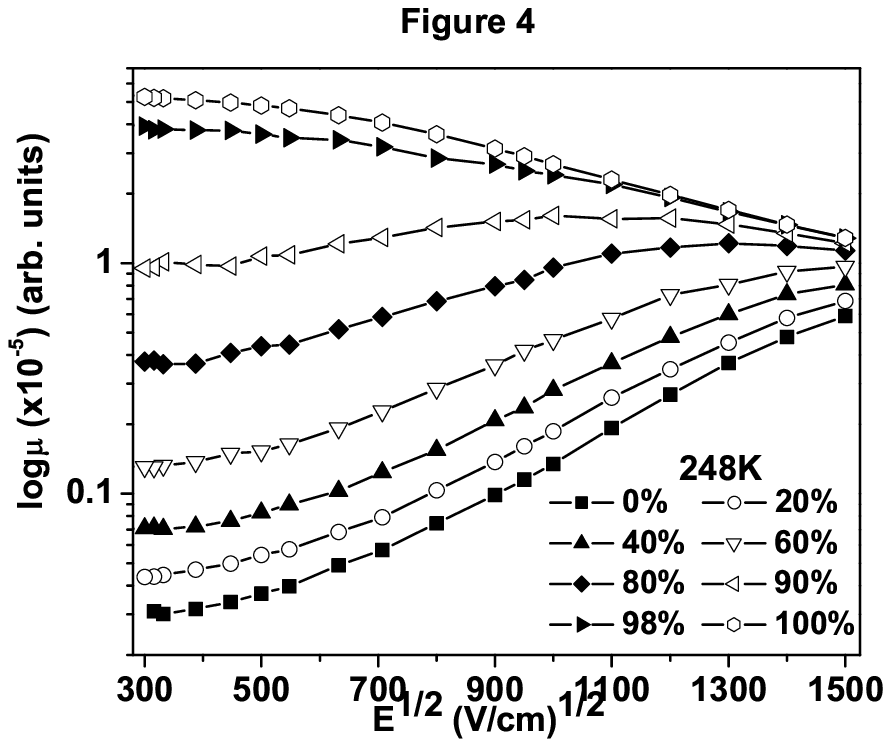}
\label{Fig4}
\end{figure}

\newpage
\begin{figure}
\includegraphics*[width=15cm]{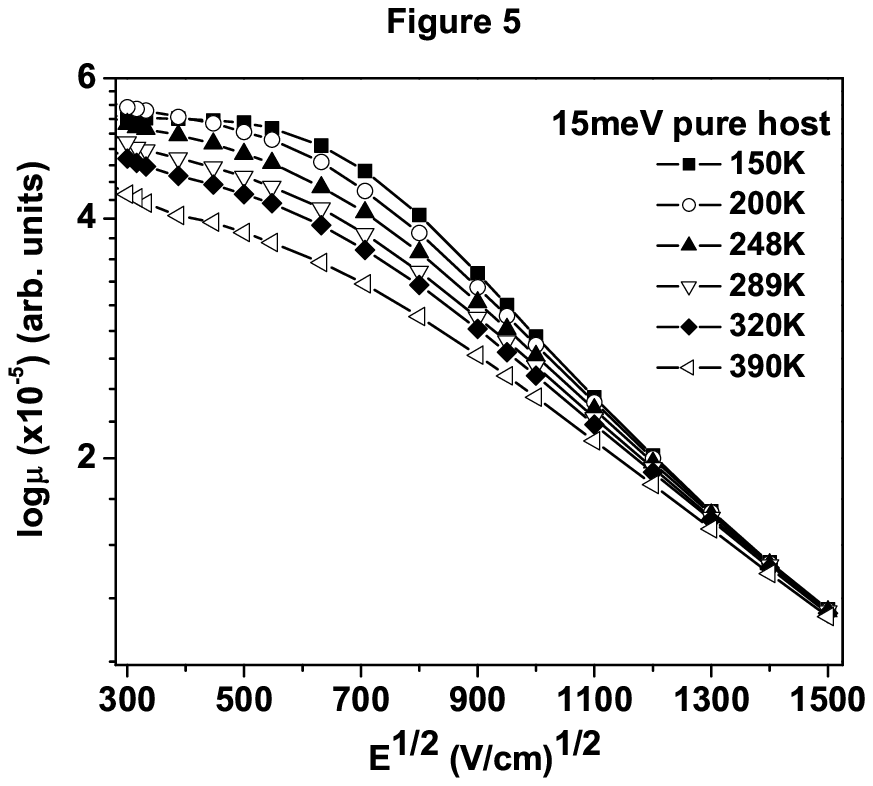}
\label{Fig5}
\end{figure}

\newpage
\begin{figure}
\includegraphics*[width=15cm]{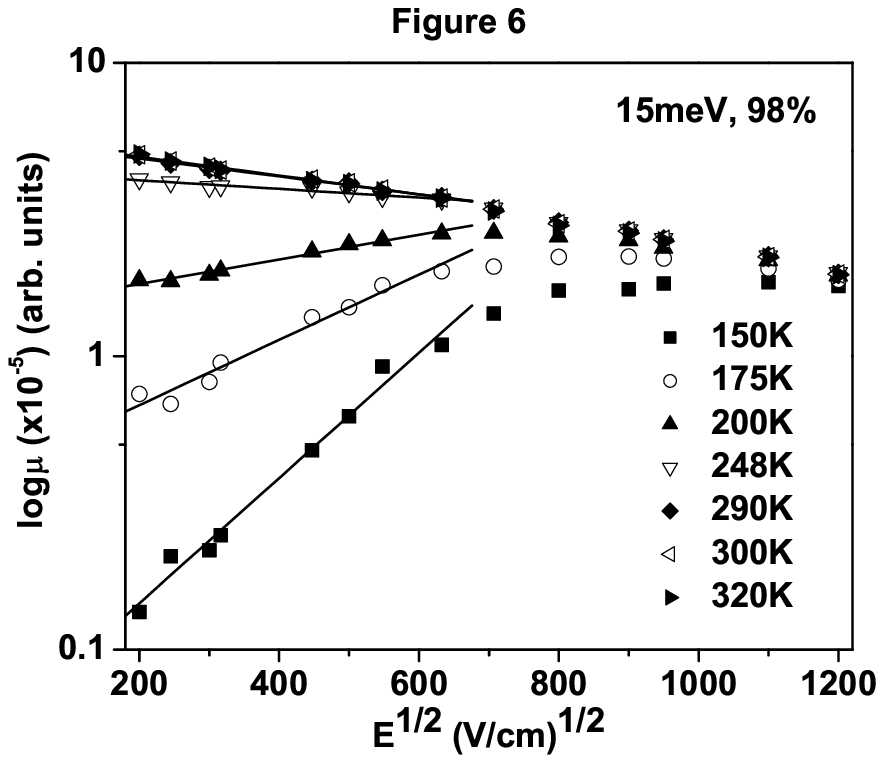}
\label{Fig6}
\end{figure}

\end{document}